\documentstyle[12pt]{article}
\begin{document}
\centerline{\large{\bf{Nonperturbative solution of the Nonconfining Schwinger Model}}}
\centerline{\large{\bf{with a generalized regularization}}}
\begin{center} Anisur Rahaman\\ Durgapur Govt. College\\ Durgapur - 14 
\\Burdwan \\West Bengal \\ India
\\e-mail: anisur@tnp.saha.ernet.in \end{center} 
\vspace{2cm}
\centerline{Abstract}
Nonconfining Schwinger model \cite{AR} is studied with a one parameter class
of kinetic energy like 
regularization. It may be thought of as a generalization over the regularization considered in \cite{AR}. 
Phasespace
structure has been determined in this new situation. The mass of the gauge boson acquires a generalized 
expression with the bare coupling constant and the parameters involved in the regularization. Deconfinement 
scenario has become transparent at the quark-
antiquark potential level.
\newpage

\section{\bf{Introduction}}

QED in $(1+1)$ dimension, e.g., Schwinger model \cite{SCH} is a very interesting exactly solvable field
theoretical model. It has been widely studied over the years by several authors in connection with the 
confinement aspect of quark \cite{LO, COL}. Here 
massless quarks interact with the Abelian gauge field. Gauge field acquires mass via a kind of dynamical 
symmetry breaking and the quarks disappear from the physical spectra. Recently, a new generalized version 
of this model hae been proposed in \cite{AR}. In that paper we find that a one parameter class of 
regularization commonly 
used to  study of chiral Schwinger model has been introduced in the vector Schwinger model. For a 
specific choice,
 i.e., for the vanishing value of the parameter the model reduces to the usual vector Schwinger model but for
 the other admissible value of this parameter the phasespace structure as well as  the the physical spectra 
gets altered remarkably. This new regularization leads to a change in the confinement scenario of the quark 
too. In fact, the quarks  gets liberated as it was happened in the Chiral Schwinger model \cite{JR}. 

In QED, a regularization gets involved when one calculates the effective action by integrating the fermions 
out. The ambiguity in the regularization has been exploited by different authors in different times and 
different interesting results have been extracted out \cite{JR, GIR, GIR1,PM, KH}. The most 
remarkable one is the chiral Schwinger model as studied by Jackiw and Rajaraman \cite{JR}. They saved the 
long suffering from the nonunitarity problems of the chiral generalization of the Schwinger model as studied 
in \cite{HAG} introducing the one parameter 
class of regularization. Regularization  plays a crucial role in the confining aspect of the 
fermions too \cite {AR, PM}. In this connection, it is fair to admit that the relation between regularization and confinement
is not yet so transparent.

In \cite{AG}, the authors showed that the regularization of the Schwinger model allows some extra flexibility
in the effective action of the theory and studied the model with a parameter dependent kinetic energy term 
for the gauge fields and got some interesting and acceptable result. We are interested here to investigate 
the Nonconfining Schwinger model with a one parameter class of kinetic energy term for the gauge field. We 
should mention here that regularization here also allows the same type of flexibility.

With this new modification the nature of the solution 
though does not change as it was happened  in \cite{AG}, but the physical mass and the bare coupling constant 
acquires a generalized form along with the two regularization parameters. This generalization allows one to 
make the physical mass zero with an unusual 
limit of one of the parameter involved in the kinetic energy term of the gauge field. This limit makes the 
deconfinement scenario more transparent when it is investigated studying the quark-antiquark potential. We 
should admit here that though there were free chiral fermion in the spectrum of the Nonconfining Schwinger model \cite{AR}, the 
deconfinement of fermions were not transparent at the quark-antiquark potential level. We therefore get 
motivated to study the deconfinement scenario in the present situation calculating the quark-antiquark
potential.

In Sec.2, we have shown how the regularization of the Schwinger model allows the flexibility to modify the 
effective action with the introduction of two parameter dependent terms. One of these two is kinetic energy
like term and the other one is conventional masslike term for the gauge fields.
Infact, it involves a generalization of the
 gauge current. It is shown that the usual fermionic operator of the Schwinger model allows the current to
 be constructed in such a way that this generalized expression is obtained.

In Sec.3, the constraint analysis of the modified bosonized effective Lagrangian has been carried out and 
the phasespace structure has been determined. Here it is shown that the physical mass and the bare coupling 
constant acquires a generalized relation.

In Sec.4, Quark-antiquark potential has been calculated for this new situation. We have shown that the 
deconfinement scenario has become transparent at the quark-antiquark potential level.

\section{\bf {Kinetic energy like regularization}}

Schwinger model is defined by the lagrangian density 
\begin{equation}
{\cal L}_F = {\bar\psi}(i\partial\!\!\!/ - eA\!\!\!/)\psi - {1\over 4} F_{\mu\nu}F^{\mu\nu}.
\end{equation}
where the Lorentz indices takes the value 0 and 1 in $(1+1)$ dimensional spacetime and the other notations 
are all standard. The coupling constant e has the dimension of mass. In $(1+1)$ dimension, the most general 
ansatz for $A_\mu$ is

\begin{equation}
A_{\nu} = -{\sqrt{\pi}\over e} (\tilde\partial_{\mu}\sigma + \partial_\mu\tilde\eta),
\end{equation}
where $\tilde\partial_\mu = \epsilon_{\mu\nu}\partial^\nu$ with $\epsilon_{01} = 1$. $\sigma$ and 
$\eta$ are two scalar fields. 
It is constructive to define a pseudo scalar field $\eta$ defined by the 
equation $\tilde\partial\eta = \partial \tilde\eta$. From the above definitions 
one  finds

\begin{equation}
F_{\mu\nu} = -{\sqrt\pi\over e}(\partial_\mu\tilde\partial_\nu - \tilde\partial_\mu\partial_\nu)\sigma = 
{\sqrt\pi\over e}\epsilon_{\mu\nu}\Box\sigma.
\end{equation}
The equation of motion to be solved here is
\begin{equation}
[i\partial\!\!\!/ + \sqrt\pi\tilde\gamma_\mu(\sigma +\eta)]\psi = 0. \label{EOST}
\end{equation}
Equation(\ref{EOST}) has the solution
\begin{equation}
\psi(x) = :e^{\sqrt\pi\gamma_5(\sigma(x) + \eta(x))}:\psi_{(0)}(x), \label{SOP}
\end{equation}
where the notations ':' indicates the normal ordering with respect to the Fock space operator 
$\sigma$ and $\eta$ and $\psi^{(0)}$ is the zero mass free Dirac field defined by
\begin{equation}
i\partial\!\!\!/ \psi^{(0)}(x) = 0. \end{equation}

Here the gauge current has been calculated using point splitting regularization. In this paper, we have made 
a generalization over the conventional construction \cite{SCH}. Current is regularized in the following way

\begin{equation}
J_\mu^{reg} = [\bar\psi(x+\epsilon)\gamma_\mu:e^{ ie\int_x^{x+\epsilon}dx^\mu[aA_\mu(x) - 
2\alpha\partial^\nu F_{\mu\nu}]}:\psi(x) -V.E.V],
\end{equation}
where $a$ and $\alpha$ are two arbitrary parameters. The term V.E.V. stands for vacuum expectation value. 
In our 
generalization two new parameters are involved. One of these two is connected with conventional masslike
regularization and the other one is connected with the kinetic energy like regularization.
This generalization though does not maintain gauge invariance like the usual vector Schwinger model it 
maintains the Lorentz invariance which is the basic need of a theory to be physically sensible. Here we should 
admit that a gauge invariant regularization some times scores over the gauge non-invariant regularization 
because it corresponds to the increased symmetry of the theory but it has to be remembered that the increased
 symmetry is not the symmetry of the physical states its mere the symmetry of the effective action which has 
to be broken by gauge fixing. Therefore, gauge non-invariant regularization should be equally acceptable. 
Moreover, the present modification is describing a generic situation.

The calculation of $J_\mu^{reg}$ is straightforward. Using $A_\mu$ and $\psi$ and keeping the terms up to 
first 
order we obtain

\begin{eqnarray}
J_\mu^{reg} &=& J_\mu^f
 - i\sqrt\pi\lim_{\epsilon\to 0} <0|\bar \psi^{(0)}(x+\epsilon)\gamma_\mu 
[a(\gamma_5\epsilon^\nu\partial_\nu +\epsilon^\nu\tilde\partial_\nu)(\sigma +\eta) 
\nonumber \\ 
&+& 2\alpha\epsilon^\nu\tilde\partial_\nu\Box\sigma]\psi^{(0)}|0>,  \end{eqnarray}
After a little simplification we have

\begin{equation}
J_\mu^{reg} = J_\mu^f - {1\over \sqrt\pi}[a{{\epsilon_\mu\epsilon_\nu 
- \tilde\epsilon_\mu\tilde\epsilon_\nu}\over {\epsilon_2}}\tilde\partial_\nu(\sigma +\eta) + 
2\alpha{{\epsilon_\mu\epsilon_\nu}\over {\epsilon^2}}\tilde\partial_\nu\Box\sigma]
,\end{equation}
where 

\begin{equation} J_\mu^f = :{\bar\psi}^{(0)}(x)\gamma_\mu\psi^{(0)}(x).\end{equation}
 Here we have used the 
identity
\begin{equation}
<0|{\bar\psi}_\alpha^{(0)}(x+\epsilon)\psi_\beta^{(0)}(x)|0> = 
-i{(\epsilon^\mu\gamma_\mu)_{\alpha\beta}\over {2\pi\epsilon^2}}\end{equation}
Taking the symmetric limit, i.e., averaging over the point splitting direction $\epsilon$ we obtain the 
final expression of $J_\mu^{reg}$:

\begin{equation}
J_\mu^{reg} = -{1\over {\sqrt\pi}}\tilde\partial_\mu(\phi+a(\sigma+\eta) + \alpha\Box\sigma)
,\end{equation}
Where $\phi$ is the potential of the free fermionic current defined by 

\begin{equation}
J_\mu^f(x) = :{\bar\psi}^{(0)}(x)\gamma_\mu\psi^{(0)}(x):, \end{equation}
with $\phi$ satisfying the bosonic equivalent of the free fermion $\psi^{(0)}$. We, therefore, have

\begin{equation}
J_\mu^{reg} (x) = -{1\over {\sqrt\pi}} \tilde\partial_\mu\phi + {e\over {\sqrt\pi}}aA_\mu - 
\alpha{e^2 \over {\sqrt\pi}}\partial^\nu F_{\mu\nu} \label{CUR}.
 \end{equation}

\section{\bf{Determination of phasespace structure from the bosonized version 
of the theory}} 

The effective bosonized lagrangian density which gives the current given in equation (\ref{CUR}) is 

\begin{equation}
{\cal L}_B = {1\over 2}\partial_\mu\phi \partial^\mu\phi - g\tilde\partial_\mu\phi A^\mu +
{1\over 2}ag^2A_{\mu}A^{\mu}
- {\tilde\alpha\over 4}F_{\mu\nu}F^{\mu\nu}. \label{BLD} 
\end{equation}
$g$ and $\tilde\alpha$ are defined by $g={e\over \sqrt\pi}$ and ${e^2\alpha\over {4\pi}} = {\tilde\alpha\over 4}$ 
for later convenience. The usual kinetic energy term for the gauge field is absorbed within the parameter
$\alpha$  Let us now proceed with the constraint analysis of the system. 
To determine the physical phasespace 
structure it is necessary to calculate the momenta corresponding to the field $A_0$, $A_1$, and $\phi$.
From the standard definition of the momentum we obtain

\begin{equation}
\pi_0=0 \label{M1},
\end{equation}
\begin{equation}
\pi_1 = F_{01}\label{M2},
\end{equation}
\begin{equation}
\pi_\phi = \dot\phi - gA_1\label{M3},
\end{equation}
where $\pi_0$,$\pi_1$ and $\pi_\phi$ are the momenta corresponding to the field 
$A_0$, $A_1$ and $\phi$. Using the equations (\ref{M1}), (\ref{M2}), and (ref{M3}), the Hamiltonian density
are calculated:

\begin{equation}
{\cal H} = {1\over 2}(\pi_\phi +gA_1)^2 + {1\over {2\tilde\alpha}}\pi_1^2 + {1\over 2}\phi'^2 + \pi_1A_0'
-gA_0\phi' - {1\over 2}ag^2(A_0^2 - A_1^2).\end{equation}
$\omega = \pi_0 = 0$, is the familiar primary constraint of the theory. The preservation of the
 constraint $\omega$
requires $[\omega(x), H(x)] = 0$, which leads to the Gauss' law as a second class constraint:

\begin{equation}
\tilde\omega = \pi_1' + g\phi' + ag^2A_1 = 0. \label{SCO}
\end{equation}
Treating (\ref{SCO}) as strong condition one can eliminate $A_0$ and obtain the reduced Hamiltonian density as
follows.

\begin{equation}
{\cal H}_r = {1\over 2}(\pi_\phi + gA_1)^2 + {1\over {2ag^2}}(\pi'_1 + g\phi')^2 + 
{1\over 2}({\pi_1^2\over {\tilde\alpha}} + \phi'^2) + {1\over 2}ag^2A_1^2.
\label{RHAM}\end{equation}
It is straightforward to show that the Dirac brackets of the remaining variable remain canonical. Using the canonical Dirac brackets the following first order equations of motion  are found out from the Hamiltonian 
density (\ref{RHAM})

\begin{equation}
\dot A_1= {1\over {\tilde\alpha}}\pi_1 -{1\over {ag^2}}(\pi_1'' + g\phi'')
,\end{equation}

\begin{equation}
\dot\phi = \pi_\phi + gA_1
,\end{equation}

\begin{equation}
\pi_\phi = {{a+1}\over a}\phi'' + {1\over {ag}}\pi_1''
,\end{equation}

\begin{equation}
\dot\pi_1 = -g\pi_\phi - (a+1)g^2A_1. \end{equation}
A little algebra converts the above first order equation into the following second order equations:

\begin{equation}
(\Box + (1+a){g^2\over {\tilde\alpha}})\pi_1 = 0, \label{SP1}
\end{equation}
\begin{equation}
\Box[\pi_1 + g(1+a)\phi] = 0. \label{SP2}
\end{equation}

Equation (\ref{SP1}) describes a massive scalar field with mass $m = \sqrt{{1+a}\over {\tilde\alpha}}g$ and 
equation (\ref{SP2}) describes a massless scalar field. Therefore, the structure of the spectra  remains 
unchanged. However, the
mass term of the massive boson gets altered and it has a crucial link with the regularization parameter 
$a$ and 
$\alpha$. In the usual vector Schwinger model mass term remains constant but it acquires a 
generalized form with the bare coupling constant $g$ in the present situation :

\begin{equation}
m^2 = ({{a+1}\over {\tilde\alpha}})g^2 .
\end{equation}
The mass here gets scaled down by a factor $\sqrt{{{a+1}\over {\tilde\alpha}}}$. The identical situation were happened in the vector meson model \cite{MES}, where mass of the physical particle gets altered because of the
variation of regularization like this situation. This is interesting to note that one can make this mass zero 
with an unusual limit $\tilde\alpha \to \infty$. The parameter $a$ can not do so because we have seen in 
\cite{AR} that $a$ has a restriction $a\ge 0$ for the theory to be physically sensible. Not only that but 
also, this unusual limit makes the  deconfining phenomena more transparent in the quark-antiquark potential 
level. We will consider this fact in our discussion in the next Section.

\section{\bf{ Calculation of quark-antiquark potential}}

In this  Section, we  calculate the quark-antiquark potential following the text \cite{ABD} and study the effect of the above mentioned 
unusual limit on the confinement aspect of quark. From the study of the previous Section we can conclude 
that the above theory can be reduced to a theory involving two free fields $\pi_1$ and $\pi_1+e\phi$. These 
two fields are not properly normalized. One can show that the properly normalized fields are

\begin{equation}
\sigma = {1\over {\sqrt{1+a}}} \pi_1, \end{equation}
\begin{equation}
\rho = {1\over {\sqrt{a(a+1)}}} [\pi_1 + g(a+1)\phi].
\end{equation} 
The original fields $\phi$ and $A_\mu$ can be expressed interms of the free fields by

\begin{equation}
\phi = {1\over {\sqrt{a+1}}} (\sqrt{a}\rho -\sigma), \end{equation}  \label{HI}
\begin{equation}
A_\mu = {1\over g} \epsilon_{\mu\nu}\partial^\nu{1\over {\sqrt{a(a+1)}}}(\sqrt{a}\sigma + \rho).\label{AI}
\end{equation}
The gauge current for this model is found out to be

\begin{equation}
J_\mu = -\epsilon_{\mu\nu}\partial^\nu\phi + agA_\mu \label{J}. \end{equation}
Substituting (\ref{HI}) and (\ref{AI}) in the equation (\ref{J}) we find

\begin{equation}
J_\mu = -\sqrt{a+1}\epsilon_{\mu\nu}\partial^\nu\sigma \label{SJ}. \end{equation}

We are now in a position to calculate the quark anti-quark potential. One can expect that the potential of
any quark-antiquark pair $(q,\bar q)$ would exhibit the behavior of the characteristic of screened charge
because of the intrinsic Higgs mechanism induced by vacuum polarization. To do that one needs to compare
the Hamiltonian in the presence of the external $(q,\bar q)$ source with that in the absence of that external
source. In the presence of the test charge $-Q$ at $-{L\over 2}$ and $+Q$ at $+{L\over 2}$ the charge density will be reduced to 

\begin{equation}
J_0 = {1\over g} Q[\delta(x+{L\over 2}) - \delta(x-{L\over 2})+\sqrt{1+a}\partial_1\sigma  = \sqrt{1+a} \partial_1(\sigma - \Theta), \end{equation}
where 
\begin{equation}
\Theta =- {g\over {\sqrt{1+a}}}Q[\theta(x+{L\over 2}) \theta({L\over 2} -x)].\end{equation}

The external charge interacts with the massive field only and change the mass of the field. Therefore, the 
presence of the external charges changes the Hamiltonian of the massive part the theory only. This change 
in the 
Hamiltonian is found as follows.
\begin{equation}
H(\Theta) = \int dx [\pi_\sigma^2 + (\partial_1\sigma)^2 + {{1+a}\over {\tilde\alpha}}g^2(\sigma -\Theta)^2]
.\end{equation} 
The potential which is the difference between the ground state energy in presence of the external charge
with the same quantity in the absence of the external charge is calculated as follows. The ground state energy in 
presence of the external source is 
\begin{equation}
E(\Theta) = <\omega_Q[\Theta]|H[\Theta]|\omega_Q[\Theta]>. \end{equation}
Using the completeness condition on the states of $\sigma$ we have
\begin{equation}
E(\Theta) = \int d\sigma d\sigma'<\omega_Q[\Theta]|\sigma><\sigma|H[\Theta]|\sigma'><\sigma'|\omega_Q[\Theta].
\end{equation}
and
\begin{eqnarray}
<\sigma|H[\Theta]|\sigma'> &=& \delta(\sigma-\sigma')\int dx [-{\partial_2\over {\partial_1\sigma(x)^2}} + 
\sigma(x)D\sigma'(x) + {g^2\over {\tilde\alpha}}(1+a) \Theta^2 \nonumber \\
&-& {2g^2\over {\tilde\alpha}} (1+a) \sigma\Theta], \end{eqnarray}
where
$D={g^2\over {\tilde\alpha}}(1+a) - \partial_1^2$.
After a little algebra the potential comes out to be 
\begin{equation}
E[\Theta] - E[0] = {1\over 2}\int[{g^2\over {\tilde \alpha}}(1+a)\Theta^2 - {g^4\over {\tilde \alpha_2}}(1+a)^2\Theta D^{-1}
\Theta]. \label{POT} \end{equation}
The operator $D^{-1}$ has the following matrix element.
\begin{eqnarray}
<x|D^{-1}|y> &=& <x|({g^2\over {\tilde\alpha}}(1+a) - \partial_1^2)^1|y> \nonumber \\
&=& {1\over 2g}\sqrt{{{1+a}\over {\tilde\alpha}}}e^{-\sqrt{{{1+a}\over {\tilde\alpha}}}}|x-y| \label{MAT}. 
\end{eqnarray}   
Substituting (\ref{MAT}) in (\ref{POT}) the quark-antiquark potential $V(L)$ is obtained:

\begin{eqnarray}
V(L) &=& E[\Theta] - E[0] \nonumber \\
&=&{{Q^2}\over {2g}}\sqrt{{{1+a}\over {\tilde\alpha}}}[1- e^{-\sqrt{{{1+a}\over {\tilde\alpha}}}gL}]
\end{eqnarray} 
$V(L)$ is the quark-antiquark potential for the  Nonconfining Schwinger model with the new regularization 
considered 
for investigation in this paper. The above potential 
comes out to be
\begin{equation}
V_{NCSM}(L) = {Q^2\over 2g}\sqrt{1+a}[1 - e^{-\sqrt{1+a}gL}], \label{POTN}
\end{equation}
for Nonconfining Schwinger model, and
\begin{equation}
V_{VSM}(L) = {Q^2\over 2g}[1 - e^{-gL}],\label{POTV}
\end{equation}
for usual vector Schwinger model.

Both in (\ref{POTN}) and (\ref{POTV}),  the potential tends to constant value for large separation  of 
the test charges, i.e., for the situation when L approaches towards infinity. This fact has good agreement
with usual vector Schwinger model since it is known that quarks gets confined during the process but the 
result was not so transparent in the analysis of the usual Nonconfining Schwinger model. It is little
 confusing too.  
There were free massless boson the spectrum and
in $(1+1)$ dimension, boson can be interpreted in terms of 
fermion. So at that stage without calculating the quark-antiquark potential it was natural to conclude 
that the fermions are not confined. However we find no signature of the presence
of fermion when we study the quark-antiquark potential carefully because only the vanishing value of the 
potential at large $L$ suggests deconfinement. Nonconfining Schwinger model with this new 
regularization can rectify the 
above confusion because if we take the limit $\tilde\alpha \to \infty$ the potential $V(L)$ approaches to
zero. Certainly, it is the signature of free quark.

\section{\bf{ Concluding remarks}}

Schwinger model was studied in \cite{AR} with a one parameter class of masslike regularization. In this 
paper,
we study this model with a more general regularization and determine its phasespace structure. The model
with this generalized  regularization remains exactly solvable. The most interesting and crucial property 
namely the deconfinement aspect of quark has been investigated calculating the quark-antiquark potential. To
modify the model we point split the current which was defined as the product of the two fermionic operator
. Schwinger gave a prescription to insert an exponential line integral of the gauge field. But this was not 
the only choice. To be more specific, it was one of the many possible choices. Following \cite{AG}, we also 
inset an extra factor in the exponential involving field strength. Of course, the introduction of 
complicated term may make the theory complicated and spoil the exactly solvable nature. However, the 
regularization chosen here keeps the solvable nature intact and makes the physical property interesting. The 
equations of motion obtained in this new situation is different from the usual vector Schwinger model 
but it is
identical in nature to the Nonconfining Schwinger model excepting the difference in the mass term of the 
massive boson. Unlike the Schwinger model, mass here acquires a generalized expression with the bare coupling constant and the parameter involved in the regularization. Certainly, the equation can be converted into free fields like the usual Vector Schwinger model. 

In the usual vector Schwinger model we find only one massive boson with a constant mass proportional to
the square of the bare coupling constant. However, in the Nonconfining Schwinger model and in the present 
situation spectra are different in nature. We find a massless boson along with the massive boson. Mass 
in both 
cases acquires a generalized expression where the regularization parameters are involved. In the Nonconfining
Schwinger model, only one regularization parameter is involved whereas in the present situation there are
two independent regularization parameter. Here the generalization is such that one can 
make the mass term zero with an unusual limit and this unusual limit can rectify the confusion of 
deconfinement scenario of the Nonconfining Schwinger model at the quark-antiquark level. In this context, we 
should mention that though there were free massless boson in the spectrum of the Nonconfining Schwinger model
the signature of deconfinement was absent in the quark-antiquark potential level. 

Question may be raised : which regularization is better gauge invariant or non-invariant? Obviously, there is 
no specific answer. We should admit that some times  gauge invariance scores over the gauge non-invariance
because of the increased symmetry of the theory. However, it has to be remember that the symmetry increased 
here is not at all a symmetry of the states its rather a symmetry of the effective action which has to be
broken by gauge fixing in order to extract out the real physical contents. So both the regularizations 
make sense
and should be acceptable in the theoretical laboratory. Concerning confinement, another question may be 
raised : which regularization
is better confining or deconfining? This question also has no specific answer. Certainly, we should make some
comments on this issue. Confinement and deconfinement is  really a crucial question. Hitherto, there is no
such model from which one can have clear explanation regarding confinement and deconfinement scenario of 
quark. Inspite of the absence of the signature of finding free quarks
from the experiment 
experimentalist of recent time have started to believe that QGP phase exists. So a practical benefit may 
follow from this work that this new version may be useful to study the 
QGP phase just as the usual vector Schwinger model has been used to study the confinement. In this sense, both
the model should be accepted in the theoretical laboratory. 

The question of confinement and deconfinement discussed in this paper is limited in $(1+1)$ dimension. It would certainly be interesting what happens to this question in $(3+1)$ dimension  since confinement and 
deconfinement are of real physical interest. Last but not the least this type of investigation may throw some
light in the fact that whether there is any connection to the confinement and regularization. More 
qualitative investigations are needed indeed in this issue.

\section{\bf{Acknowledgments}}

It is a pleasure to acknowledge Prof. P. Mitra of Saha Institute of Nuclear Physics for helpful discussion

\end{document}